\documentclass[showpacs,preprint]{revtex4}
\usepackage{graphicx}
\usepackage{subfigure}
\usepackage{amssymb}
\usepackage{amsmath}
\usepackage{bm}
\usepackage{latexsym}
\begin{document}

\title{Non-Markovian dynamical effects and time evolution of entanglement entropy of a dissipative two-state system}

\author{Zhiguo L\"{u}\footnote{Email: zglv@sjtu.edu.cn}, Hang Zheng}
\affiliation{ Department of Physics, Shanghai Jiao Tong
University, Shanghai 200240, China}

\begin{abstract}
We investigate the dynamical information exchange between a
two-state system and its environment which is measured by von
Neumann entropy. It is found that in the underdamping regime, the
entropy dynamics exhibits an extremely non-Markovian
oscillation-hump feature, in which oscillations manifest quantum
coherence and a hump of envelop demonstrates temporal memory of
bath. It indicates that the process of entropy exchange is
bidirectional. When the coupling strength increases a certain
threshold, the hump along with ripple disappears, which is
indicative of the coherent-incoherent dynamical crossover. The
long-time limit of entropy evolution reaches the ground state value
which agrees with that of numerical renormalization group.

\end{abstract}


\maketitle

The non-equilibrium evolution of open quantum systems is one of the
most challenging and intriguing problems of contemporary research in
both theoretical and experimental physics. The transient dynamics
can be harnessed and controlled desirably in quantum information
processing. The first step to manipulate it is to understand how it
evolves in a short time interval. It is known that the correlations
of open system with its surrounding environment lead to finite
lifetime of quantum superpositions, which give rise to the evolution
from pure states into mixed ones. It is often stated that
decoherence causes the system to become entangled with its
environment, and the entanglement between them can be measured
quantitatively by von Neumann entropy\cite{Vedral}. The main
questions which now arise are: How does the entropy or quantum
information flow from system to environment? After the state of
system is initialized as a pure state without entanglement, how does
the entropy evolve to its long-time limit (monotonously or not)? Is
the process of information transfer between bath and system,
unidirectional or bidirectional? In this paper, as far as we know,
it is the first time to show the time evolution of entropy for open
system which exhibits extremely non-Markovian characters and point
out the process of entropy exchange is bidirectional in the
underdamping region.

The dissipative two-state system (TSS), which is also called the
spin-boson model, as a simple paradigm of open system, is a generic
model which can be widely used to describe a large number of
physical and chemical processes, such as the defect-tunneling and
electron transfer, and applied to clarify very interesting quantum
phenomena, such as decoherence and dephasing\cite{rmp,book}. The
open system inevitably encounters decoherence which renders a
quantum superposition state to decay into a classical, statistical
mixture of states. Thus, derivation of the reduced density matrix is
a central goal in order to describe its evolution. Based on weak
coupling assumption, a Markovian master equation could give its
dynamics. However, strong interactions with low-temperature
reservoirs give rise to large system-environment correlations which
generally result in a failure of the Markovian approximation. In
this case, the system dynamics possesses long memory times and
exhibits a pronounced non-Markovian behaviors \cite{Breuer}. Thus,
it is significant to show the temporary evolution of system by a
non-Markovian approach, especially, in the case of the strong
coupling to its bath. The non-Markovian approach can investigate not
only more complicated situation where Markovian approximation is
unreachable but also different spectral densities between the system
and the environment.

The entanglement entropy has been considered in many works. They
mainly focus on the static or ground state properties and find some
important results\cite{cm,hur,kjc}. Costi and McKenzie used a
numerical renormalization group (NRG) treatment to study the entropy
of the ground state as a function of coupling\cite{cm}. Recently,
Hur and coworkers applied the NRG to study the quantum phase
transition and found that there is a cusp in the entanglement
entropy accompanying with quantum phase transition\cite{hur}. As
explicitly pointed out by Costi and McKenzie, it is fascinating to
show how entropy varies with time after the qubit is initially
prepared in a certain state without entanglement\cite{cm}. In order
to calculate the dynamics of entanglement entropy we present an
analytical approach based on a unitary transformation method without
the Markovian approximation. It works well in the parameter regime
$0<\alpha<1$ and $0<\Delta<\omega_c$, and can reproduce well known
non-perturbation results obtained by various methods, such as the
coherence-incoherence transition, which have been studied in our
previous work\cite{zheng}. The approach does not invoke the
rotating-wave approximation so as to take into account the effects
of count-rotating terms on transient dynamics. Since the quantum
manipulation can be effectively made in the coherent region, we
would give the evolution of entropy for this regime at $T=0$. We
find the non-Markovian entropy evolution with pronounced small
oscillations feature in the weak coupling, which demonstrates
quantum coherence. As the coupling increases, a hump along with
ripple clearly emerges in the short time characterizing the temporal
memory of bath, eventually the dissipative effects quench some
oscillations and the hump near the coherence-incoherence transition.

The spin-boson model reads\cite{rmp,book}
\begin{eqnarray}
&&H=-{\frac{\Delta}{2}} \sigma _{x} +\sum_{k}\omega _{k}b_{k}^{\dag }b_{k} +%
{\frac{1}{2}}\sum_{k}g_{k}(b_{k}^{\dag }+b_{k})\sigma _{z}.
\end{eqnarray}
Standard notations are used\cite{book}, $\Delta $ is the bare
tunneling matrix and $g_{k}$ the coupling constant. The effect of
environment is determined by its spectral density:
$\sum_{k}g^2_{k}\delta(\omega-\omega_k)=2\alpha\omega
\theta(\omega_c-\omega)$, where $\alpha$ is the dimensionless
coupling constant, $\omega_c$ is a cutoff frequency and $\theta(x)$
is the usual step function (In the work the spectrum is Ohmic type,
and we set $\hbar=k_{B}=1$). Although the model seems quite simple,
it is in general not exactly solvable and a large variety of
approximate analytical and numerical methods have been proposed and
implemented to study its ground state and
dynamics\cite{cm}{-}\cite{ks}.

\textbf{\textit{Unitary transformation.}}  A unitary transformation,
which is defined as $H^{\prime }=\exp (S)H\exp (-S)$, is applied to
$H$ in order to take into account the correlation between the spin
and bosons\cite{zheng,sh}.  The form of generator is proposed,
\begin{equation}
S=\sum_{k}\frac{g_{k}}{2\omega _{k}}\xi _{k}(b_{k}^{\dag
}-b_{k})\sigma _{z} ,
\end{equation}
where a $k$-dependent function
$\xi _{k}$ is introduced \cite{zheng}. The transformation can be
performed to the end and the result is $H^{\prime }=H_{0}^{\prime
}+H_{1}^{\prime }+H_{2}^{\prime }$,
\begin{eqnarray}
H_{0}^{\prime }&=&-{\frac{\Delta_r}{2}}  \sigma _{x} +\sum_{k}\omega
_{k}b_{k}^{\dag }b_{k} -\sum_{k}\frac{g_{k}^{2}}{4\omega _{k}}\xi
_{k}(2-\xi _{k}),     \\
H_{1}^{\prime }&=&{\frac{1}{2}}\sum_{k}g_{k}(1-\xi _{k})(b_{k}^{\dag
}+b_{k})\sigma _{z} -{\frac{\Delta_r}{2}} i\sigma_y B,   \\
H_{2}^{\prime }&=&-{\frac{\Delta}{2}}\sigma_x\left( \cosh \{B\}-\eta
\right)-{\frac{\Delta}{2}} i\sigma_y\left( \sinh \{B\}-\eta B
\right),
\end{eqnarray}
where $B=\sum_{k}\frac{ g_{k}}{\omega _{k}}\xi _{k}(b_{k}^{\dag
}-b_{k})$ and $\Delta_r = \eta \Delta$. The renormalizied factor of
tunneling is $\eta =\exp [-\sum_{k}\frac{g_{k}^{2}}{2\omega
_{k}^{2}}\xi _{k}^{2}]$. $H^{\prime}_0$ is the unperturbed part of
$H^{\prime}$ and, obviously, it can be solved exactly since the spin
and bosons are decoupled.  The ground state of $H_{0}^{\prime}$ is
$|g_{0}\rangle =|s_{1}\rangle |\{0_{k}\}\rangle$ ($\sigma_x
|s_{1}\rangle = |s_{1}\rangle$,$|\{0_{k}\}\rangle$ is the vacuum
state  for every boson mode $n_{k}=0$). $H_{1}^{\prime }$ and
$H_{2}^{\prime }$ are treated as perturbation and they should be as
small as possible. For this purpose $\eta$ is determined to make
$\mbox{Tr}_B(\rho_B H'_2)=0$, where $\rho_B$ is the density operator
of bath.
Besides, $\xi _{k}$ is determined as
\begin{eqnarray}
\xi _{k}=\frac{\omega _{k}}{\omega _{k}+\Delta_r },
\end{eqnarray}
and because of this form $H_1^{\prime}$ is rewritten as
\begin{eqnarray} \label{H1'}
&&H_{1}^{\prime}= \sum_{k} V_{k}\left[ b_{k}^{\dag }\sigma
_{-}+b_{k}\sigma _{+}\right],
\end{eqnarray}
where $V_k=\Delta_r g_k\xi_k/\omega_k$ and
$\sigma_{-}=(\sigma_z-i\sigma_y)/2$,
$\sigma_{+}=(\sigma_z+i\sigma_y)/2$. When $T=0$ it is easy to check
that $H_{1}^{\prime }|g_{0}\rangle =0$. This is essential in our
approach.

In our treatment $H'_0$ is treated as the unperturbed Hamiltonian,
in which the tunneling has been already renormalized by $\eta$
coming from the contribution of diagonal transition of bosons.
$H'_1$ is the perturbation relating to the non-diagonal transition
of single-boson, and $H'_2$, containing all other multi-boson
non-diagonal transitions, is omitted because its contribution to
physical quantities is $O(g^4_k)$ and higher. Note that $0 \leq
\xi_k \leq 1$. $\xi_k$ measures the adiabatic intensity of the
particle interacting with its environment\cite{zheng}. $\xi_k \sim
1$ if $\omega_k \gg \Delta_r$, while $\xi_k \ll 1$ for $\omega_k \ll
\Delta_r$. In addition, by the choice of $\xi_k$, $H'_1$ has taken
into account the effects of counter-rotating terms. In other words,
the bare coupling $g_k/2$ in the original Hamiltonian is replaced by
the renormalized coupling $ V_k $ after the unitary transformation.

\textit{\textbf{Density operator.}}  In order to show the quantum
dynamics, we would first give the density operator in
Schr\"{o}dinger representation, $\rho_{SB}(t)$ with Hamiltonian $H$,
where the subscript SB stands for the spin-boson model. For the
transformed Hamiltonian $H'$ the density operator is
$\rho'_{SB}(t)=e^S \rho_{SB}(t) e^{-S}$. The density operator in the
interaction representation is $ \rho^{\prime
I}_{SB}(t)=\exp(iH'_0t)\rho'_{SB}(t)\exp(-iH'_0t)$. By the equation
of motion for $\rho^{\prime I}_{SB}(t)$ \cite{scu}, we obtain the
master equation
\begin{eqnarray} \label{meq}
&&\frac{d}{dt}\rho^{\prime I}_{S}(t)
=-\int^t_0\mbox{Tr$_B$}[H'_1(t),[H'_1(t'),\rho^{\prime
I}_S(t')\rho_B]]dt'.
\end{eqnarray}
where $\rho^{\prime I}_{S}(t)=\mbox{Tr$_B$}\rho^{\prime I}_{SB}(t)$
and $H'_1(t)=\exp(iH'_0t) H'_1 \exp(-iH'_0t)$. It is known that one
can arrive at the Born-Markov approximation equation neglecting
retardation in the integration, i.e., $\rho^{\prime I}_S(t')$ is
replaced by $\rho^{\prime I}_S(t)$. Our treatment is beyond this
approximation.

At $t=0$, the usual initial density operator is
$\rho_{SB}(0)=\rho_S(0)\rho_B=\left(
\begin{array}{cc}
1 & 0\\
0 & 0
\end{array}
\right) \rho_B$. Then we can get the initial condition for our
calculations: $\rho^{\prime
I}_{SB}(0)=\rho'_{SB}(0)=e^S\rho_{SB}(0)e^{-S}$ leads to $
\rho^{\prime I}_S(0)=\left(
\begin{array}{cc}
1 & 0\\
0 & 0
\end{array}
\right). $ The calculation is up to the second order $g^2_k$ and and
the details are shown in the Appendix. The solution of reduced
density operator $\rho^{\prime}_S(t)=\left(
\begin{array}{cc}
\rho^{\prime}_{11} & \rho^{\prime}_{12}\\
\rho^{\prime}_{21} & \rho^{\prime}_{22}
\end{array}
\right)$ is
\begin{eqnarray}\label{pt}
\rho^{\prime}_{11}(t)-\rho^{\prime}_{22}(t)=\frac{1}{4\pi
i}\int^{-\infty}_{\infty} e^{-i\omega t}d\omega F^*(\omega)
+\frac{1}{4\pi i}\int^{\infty}_{-\infty} e^{i\omega t}d\omega
F(\omega),
\end{eqnarray}
\begin{eqnarray}
&&\rho^{\prime}_{12}(t)+\rho^{\prime}_{21}(t)=1-\frac{1}{2\pi
i}\int^{\infty}_{-\infty} \frac{e^{i\omega t} d\omega}
{\omega-\sum_k\left[\frac{ V_k^2}{\omega+\omega_k-\Delta_r-i0^{+}}
+\frac{ V_k^2}{\omega-\omega_k+\Delta_r-i0^{+}}\right]},
\end{eqnarray}
where $ F(\omega)=(\omega-\Delta_r-\sum_k\frac{
V_k^2}{\omega-\omega_k-i0^+})^{-1}. $ The real and imaginary parts
of $ \sum_{k} V_{k}^2/(\omega -i0^{+}-\omega _{k})$ are denoted as
\begin{eqnarray}
&&R(\omega )=-2\alpha \frac{\Delta_r ^{2}}{\omega +\Delta_r }
\left\{ \frac{\omega _{c}}{\omega _{c}+\Delta_r }-\frac{\omega }{
\omega +\Delta_r }\ln \left[ \frac{\omega (\omega _{c}+\Delta_r
)}{\Delta_r
(\omega _{c}-\omega )}\right] \right\}, \\
&&\gamma (\omega )=2\alpha \pi \omega \frac{  \Delta_r^{2}}{(\omega
+\Delta_r )^{2}} (0\le\omega\le\omega_c),
\end{eqnarray}
respectively.

\textit{\textbf{Dynamical quantities.}}  In what follows we
calculate the dynamical quantities,
$\langle\sigma_{\tau}(t)\rangle=\mbox{Tr}_S\mbox{Tr}_B[\rho_{SB}\sigma_{\tau}](\tau=x,y,z)$.
The reduced density operator of the original Hamiltonian $H$ is
$\rho_{S}(t)=\mbox{Tr$_B$}\rho_{SB}(t)$, which can be expressed as
$\rho_S(t)={\frac{1}{2}}[1+\sum_{\tau}\langle\sigma_{\tau}(t)\rangle\sigma_{\tau}].$
First, we calculate $\langle\sigma_z(t)\rangle$ which is usually
denoted as $P(t)$ in the literature,
\begin{eqnarray} \label{int-pt}
&&P(t)=\mbox{Tr}_S\mbox{Tr}_B(\rho_{SB}(t)\sigma_z)
=\frac{1}{\pi}\int^{\omega_c}_{0}
d\omega\frac{\gamma(\omega)\cos(\omega
t)}{[\omega-\Delta_r-R(\omega)]^2+\gamma^2(\omega)},
\end{eqnarray}
since $\mbox{Tr$_B$}\rho_B=1$. The integration in Eq.(\ref{pt}) can
be done approximately by the residue theorem, $P(t)=\cos(\omega_0
t)\exp(-\gamma t)$, where $\omega_0$ is the solution of equation
$\omega_0-\Delta_r-R(\omega_0)=0$, and $\gamma$ is the
Wigner-Weisskopf approximation of $\gamma(\omega)$: $\gamma ={\frac{
\pi}{2}}\alpha \Delta_r$. The solution $\omega_0$ is real only when
$\alpha<\alpha_c$, $ \alpha_c=(1+\Delta_r/\omega_c)/2$. It becomes
the well-known result $\alpha_c=1/2$ in the scaling limit $
\Delta/\omega_c\ll 1$\cite{rmp,book}. For $\alpha >\alpha _{c}$
there is no real solution $\omega _{0}$ and it means that $\alpha
=\alpha _{c}$ determines the critical point corresponding to the
coherent-incoherent transition. The coherent regime $\alpha <\alpha
_{c}$ can be divided into the underdamping part and the overdamping
one by a criterion $\omega_0>\gamma(\omega_0)\mbox{~(underdamping),
or~}\omega_0<\gamma(\omega_0)\mbox{~(overdamping).} $ In the scaling
limit $\Delta/\omega_c \ll 1$ the point where
$\omega_0=\gamma(\omega_0)$ is at $\alpha^{*}_c=0.325$, which is
very close to previous results $\alpha=1/3$ or
$0.3$\cite{cos,bulla}. From Eq. (1) one can get a relation between
$\langle\sigma_y(t)\rangle$ and $\langle\sigma_z(t)\rangle$,
$\langle\sigma_y(t)\rangle=-\frac{1}{\Delta}\frac{d}{dt}\langle\sigma_z(t)\rangle$,
since $i[H,\sigma_z]=-\Delta\sigma_y$. $\langle\sigma_x(t)\rangle$
can be calculated in the following,
\begin{eqnarray}\label{xt}
&&\langle\sigma_x(t)\rangle=\mbox{Tr}_S\mbox{Tr}_B(\rho^{\prime}_{SB}(t)e^{S}\sigma_xe^{-S})
=\eta \left\{1-\frac{1}{\pi}\int^{\infty}_{-\infty}
\frac{\Gamma(\omega)\cos(\omega t)d\omega}{[\omega-\Sigma(\omega)]^2
+\Gamma^2(\omega) }\right\},
\end{eqnarray}
where $\Sigma(\omega)=R(\Delta_r+\omega)-R(\Delta_r-\omega)$ and
$\Gamma(\omega)=\gamma(\Delta_r+\omega)+\gamma(\Delta_r-\omega)$.
One can check that the initial conditions
$\langle\sigma_x(0)\rangle=0, \langle\sigma_y(0)\rangle=0,
\langle\sigma_z(0)\rangle=1 $ are exactly satisfied. Besides, $
\langle\sigma_x(\infty)\rangle=\eta$,
$\langle\sigma_y(\infty)\rangle=0$,
$\langle\sigma_z(\infty)\rangle=0$, which are the correct results
for thermodynamic equilibrium state\cite{cm}.

\textit{\textbf{Entropy of entanglement.}} The entropy
(indeterminacy of the state) is a measure of the missing information
compared with the pure state of the composite system. The more lost
information about the composite state, the more information is
contained in the correlation between the substates. The greater the
entropy of system, the more strongly is the pure state of composite
system correlated and thus entangled\cite{aud}.  To see what happens
to the coherence properties due to the interaction between the
system and its surrounding starting from a pure state, we use the
von Neumann entropy. It is defined as  $S(t) = -\mbox{Tr} (\rho_S
\log_2 \rho_S)$, which is a measure of the entanglement between
them. It may be expressed in terms of the eigenvalues $\lambda_{\pm}
(t)= 1/2\pm
\sqrt{\langle\sigma_x(t)\rangle^2+\langle\sigma_y(t)\rangle^2+\langle\sigma_z(t)\rangle^2}/2$
of the density operator $\rho_S$ as, $ S(t) = -\lambda_{+}\log_2
\lambda_{+} -\lambda_{-} \log_2\lambda_{-}.$

Form the Hamiltonian (Eq. 1), it predicts that
$\langle\sigma_y(\infty)\rangle=0$,
$\langle\sigma_z(\infty)\rangle=0$, and only
$\langle\sigma_x(\infty)\rangle$ is nonvanishing in the delocalized
phase, which verify our obtained results. So, the entropy in
long-time limit $S_{eq}$ is given by
$\lambda_{\pm}=1/2\pm\langle\sigma_x\rangle/2$, which is shown in
Fig. 1 along with the NRG results\cite{cm}. As $\alpha$ increases,
$S_{eq}$ becomes large. When $\alpha\rightarrow 1$, $S_{eq}$ tends
to one. In the scaling limit, for $\alpha > 1$,
$\langle\sigma_x\rangle=0$ and the system remains its initial state,
thus $|\langle\sigma_z\rangle|=1$ and $S_{eq}=0$. In other words,
the transition between localized and delocalized phase occurs at
$\alpha=1$ and the entropy decreases from unity to zero abruptly.


In order to calculate the entropy, $\langle\sigma_x(\infty)\rangle$
can also be evaluated using the NRG applied to the equivalent
anisotropic Kondo model. The NRG data shown in Fig.1 are taken from
Ref.\cite{cm}.  It is seen that for small tunneling our result is in
good agreement with those of NRG. However, with increasing large
tunneling some discrepancies appear for moderate values of the
coupling. We think that it comes possibly from the NRG
discretization\cite{bulla}.

The dynamics of entanglement entropy displays extremely
non-Markovian features. Figure 2 shows $S(t)$ for different
couplings with $\Delta= 0.1\omega_c$. For $\alpha < \alpha_c^{*}$,
the entropy increases non-monotonically from zero to a finite value
($S_{eq}$) with explicit oscillations, and would not come close to
saturation in the short-time interval which means that quantum
coherence is not directly destroyed by the bath. At the same time,
the envelope of entropy exhibits a hump characterizing short-time
memory of bath. On the other hand, the oscillation-hump feature
demonstrates that the process of entropy exchange is bidirectional.
To better understand the nature of the oscillation-hump feature and
the large contribution from quantum fluctuations, we should consider
the elements of the reduced density matrix.
$\langle\sigma_y(t)\rangle$ and $\langle\sigma_z(t)\rangle$ exhibit
oscillations which represent coherence. So, the dominant
contribution to the oscillatory signal comes from
$\langle\sigma_y(t)\rangle ^2+\langle\sigma_z(t)\rangle^2$, while
the trend of entropy evolution ascribes to
$\langle\sigma_x(t)\rangle$. Thus, the oscillation of entropy shows
the coherent evolution in coherent regime.

%

As coupling increases, oscillations become obviously weaker with
small amplitudes and the envelop of entropy rises rapidly with
small hump(To see Fig. 2b) due to the effects of strong
dissipation. Near the crossover from coherent to incoherent regime
$\alpha \sim 1/2$, entropy shape displays faster rising behaviors
without oscillation and the hump disappears, which is an important
character corresponding to the coherent-incoherent crossover. Note
that $S_{eq}$ is analytic and continuous at $\alpha_c$ because no
phase transition happens at this point while the dynamical
crossover from damped oscillatory to pure decaying behaviors takes
place. Thus, we can not extract a distinguishable feature of this
crossover from $S_{eq}$ because of its character of
thermodynamical equilibrium even if $\alpha > \alpha_c$ $S_{eq}$
is near to its saturation. Therefore, only transitory dynamics of
entropy could give the indicator of the crossover even though
$S_{eq}$ might also be regarded as an interesting order parameter
to mark quantum phase transitions.

The evolution of entanglement entropy is very different from that of
Markovian approximation. The dynamics of $S(t)$ is shown in Fig. 3a
with several tunnelings for $\alpha=0.2$ as well as the
corresponding Markovian results. In the Markovian evolution, the
system undergoes a smooth and fast relaxation to its final
statistical mixture. It is found that there is no short-time
oscillations in the Markovian evolution. Thus, the transient
oscillatory behaviors of entropy dynamics can not be correctly
described by the Markovian approximation. Nevertheless, in the
long-time limit, Markovian results are consistent with $S_{eq}$ as
expected. The oscillation of the entropy is a hallmark of
non-Markovian dynamics in the coherent regime which is unexpected in
the Markovian dynamics. From the scaled entropy $S(t)/S_{eq}$ in
Fig. 3b, one can see that the entropy displays almost synchronously
with different amplitudes of oscillations for any tunneling and
eventually goes wiggly down to $S_{eq}$. It indicates that the
system exchanges entropy frequently with its environment in the
short time. In this case, the oscillations are more pronounced for
the enhancement of coherence involved transition between two states.
From another point of view, the ability of exchanging information
becomes strong for the system with increasing tunneling and it
remains coherence for a longer time. (Note that the unit of time is
$\eta \Delta$, which becomes explicitly larger with increasing
tunneling.)

In the coherent regime, a sufficient number of quantum manipulations
can be performed within the coherent time. The need to maintain
quantum coherence during the operation is especially difficult to
achieve in solid state systems such as quantum dots which couple
relatively strongly to uncontrollable environmental degrees of
freedom, leading to decoherence.  Only in the underdamping regime,
the quantum control has more efficiency. The promising experimental
proposal that entanglement entropy can be measured in Cooper pair
box or quantum dot scheme is suggested by Kopp and Hur
recently\cite{hur}. We really expect that experimental setup is
capable of testing our predictions and such measurements would
provide a proof of the existence of oscillations in the entropy
evolution although it is not easy to probe small signals in the
background of noises and thermal fluctuations.

\textit{\textbf{Summary.}}  The entanglement entropy dynamics of
dissipative TSS is studied by means of the analytical approach on
the basis of a unitary transformation. Analytical results of the
quantum dynamics, described by the $\rho_S(t)$, is obtained for the
general finite $\Delta /\omega _{c}$ case. The entanglement entropy
evolution from a pure state is shown with explicit non-Markovian
features. Our approach is quite simple and tractable without
spectral structure dependence, and it could trigger many future
applications in other more complicated coupling systems with
realistic spectrum function, such as superconducting qubit with
Lorentz spectrum.

Here are a few words about the key ingredient of the approach. The
purpose of our unitary transformation is to find a better way to
divide the transformed Hamiltonian into unperturbed part
$H^{\prime}_0$, which can be treated exactly, and perturbation ones
$H^{\prime}_1+H^{\prime}_2$, which may be treated by perturbation
theory. In $H'_0$ the tunnelling has been already renormalized by
$\eta$ which comes from the contribution of diagonal transition of
bosons. $H'_1$ is related to the non-diagonal transition of
single-boson and all other multi-boson non-diagonal transitions are
contained in $H'_2$. If one treats the coupling term in the original
Hamiltonian $H$ as the perturbation, the dimensionless expanding
parameter is $g_{k}^{2}/\omega _{k}^{2}$. For Ohmic bath $s=1$ it is
$2\alpha/\omega $ which is logarithmic divergent in the infrared
limit. By choosing the form of $\eta $ and introducing the function
$\xi_k$ in the unitary transformation it is possible to treat
$H_{1}^{\prime }$ and $H_{2}^{\prime }$ as perturbation because of
the following reason. On account of the form of $\eta$
$H_{2}^{\prime}$ can be treated as perturbation because its
contribution is zero at second order of $g_k$. The effect of the
coupling term in $H^{\prime}$ ($ H^{\prime}_1$) can be safely
treated by perturbation theory because the infrared divergence in
the original perturbation treatment for $H$ is eliminated by making
choice of the function form $\xi_k$. The expanding parameter ($s=1$)
is $g_{l}^{2}\xi _{l}^{2}/\omega _{l}^{2}\sim 2\alpha\omega /(\omega
+\eta \Delta )^{2}$, which is finite in the infrared limit. Besides,
our approach is well checked not only by the initial values of the
correlation functions and entanglement entropy, such as $P(t=0)=0$,
$S(t=0)=0$, and their long time limits such as $P(\infty)=0$,
$S(\infty)=S_{eq}$.

This work was supported by the China National Natural Science
Foundation (Grants Nos. 10734020 and 90503007).

\section*{\textbf{Appendix}}

\setcounter{equation}{0}
\renewcommand{\theequation}{A\arabic{equation}}

In this Appendix we list the details of solving the master equation
(8). The integration in Eq.(8) can be done as follows,
\begin{eqnarray}
&&-\int^t_0\mbox{Tr$_B$}[H'_1(t),[H'_1(t'),\rho^{\prime
I}_S(t')\rho_B]]dt'\nonumber\\
&&=-\sum_kV^2_k\int^t_0dt'\left\{\left[n_k\sigma_{-}\sigma_+\rho^{\prime
I}_S(t')-(n_k+1)\sigma_{-}\rho^{\prime
I}_S(t')\sigma_{+}\right.\right.\nonumber\\
&&\left.-n_k\sigma_{+}\rho^{\prime
I}_S(t')\sigma_{-}+(n_k+1)\rho^{\prime
I}_S(t')\sigma_{+}\sigma_{-}\right]\exp[i(\omega_k-\Delta_r)(t-t')]\nonumber\\
&&+\left[(n_k+1)\sigma_{+}\sigma_{-}\rho^{\prime
I}_S(t')-n_k\sigma_{+}\rho^{\prime I}_S(t')\sigma_{-}
-(n_k+1)\sigma_{-}\rho^{\prime
I}_S(t')\sigma_{+}\right.\nonumber\\
&&\left.\left.+n_k\rho^{\prime
I}_S(t')\sigma_{-}\sigma_{+}\right]\exp[-i(\omega_k-\Delta_r)(t-t')]\right\},
\end{eqnarray}
where $n_k=1/[\exp(\beta\omega_k)-1]$ is the Bose function. Thus,
Eq.(8) can be solved by the Laplace transformation. If we denote
\[
\rho^{\prime I}_S(p)=\left(
\begin{array}{cc}
\rho^{\prime I}_{11} & \rho^{\prime I}_{12}\\
\rho^{\prime I}_{21} & \rho^{\prime I}_{22}
\end{array}
\right),
\]
the solution of Eq.(A1) is
\begin{eqnarray}
&&\rho^{\prime I}_{11}-\rho^{\prime I}_{22}=
\frac{1/2}{p+\sum_k\frac{
V_k^2\coth(\omega_k/2T)}{p+i(\omega_k-\Delta_r)}}+\frac{1/2}{p+\sum_k\frac{
V_k^2\coth(\omega_k/2T)}{p-i(\omega_k-\Delta_r)}},\\
&&\rho^{\prime I}_{12}-\rho^{\prime I}_{21}=
\frac{1/2}{p+\sum_k\frac{
V_k^2\coth(\omega_k/2T)}{p+i(\omega_k-\Delta_r)}}-\frac{1/2}{p+\sum_k\frac{
V_k^2\coth(\omega_k/2T)}{p-i(\omega_k-\Delta_r)}},\\
&&\rho^{\prime I}_{12}+\rho^{\prime I}_{21}=\frac{\sum_k
\frac{2V^2_k}{p^2+(\omega_k-\Delta_r)^2}}{p\left(1+2\sum_k\frac{
V_k^2\coth(\omega_k/2T)}{p^2+(\omega_k-\Delta_r)^2}\right)}.
\end{eqnarray}
Using the relation between Schr\"{o}edinger and interaction
representation and making the Laplace inverse-transformation, we can
get
\begin{eqnarray}
&&\rho^{\prime}_{11}(t)-\rho^{\prime}_{22}(t)=\cos(\Delta_r
t)(\rho^{\prime I}_{11}(t)-\rho^{\prime I}_{22}(t))-i\sin(\Delta_r
t)(\rho^{\prime
I}_{12}(t)-\rho^{\prime I}_{21}(t))\nonumber\\
&&=\frac{1}{4\pi i}\int e^{pt}dp\left
\{\frac{1}{p+i\Delta_r+\sum_k\frac{
V_k^2\coth(\omega_k/2T)}{p+i\omega_k}}+\frac{1}{p-i\Delta_r+\sum_k\frac{
V_k^2\coth(\omega_k/2T)}{p-i\omega_k}}\right\},\nonumber\\
\\
&&\rho^{\prime}_{12}(t)+\rho^{\prime}_{21}(t)=\rho^{\prime
I}_{12}(t)+\rho^{\prime I}_{21}(t)=\frac{1}{2\pi i}\int
e^{pt}dp\frac{\sum_k
V^2_k\frac{2}{p^2+(\omega_k-\eta\Delta)^2}}{p\left(1+2\sum_k\frac{
V_k^2\coth(\omega_k/2T)}{p^2+(\omega_k-\eta\Delta)^2}\right)}.
\end{eqnarray}
The integration path is on a line parallel to the imaginary axis of
complex $p$ plane from $p=0^{+}-i\infty$ to $p=0^{+}+i\infty$.

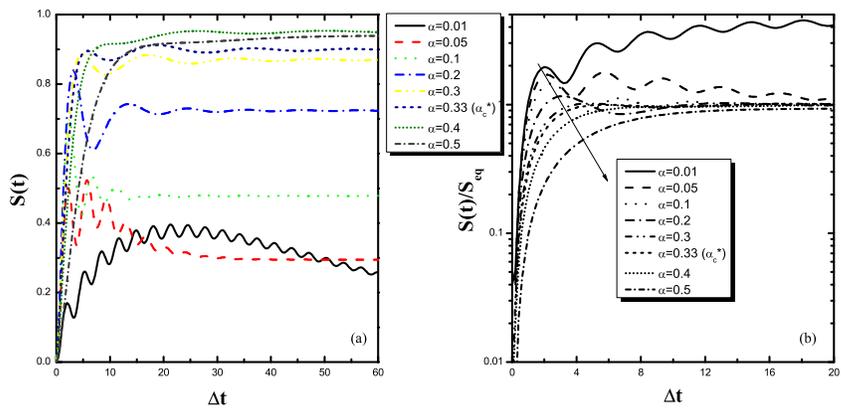
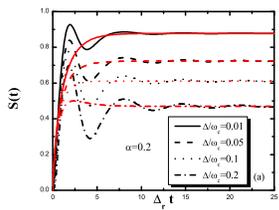
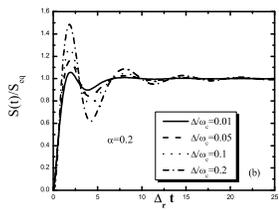
\end{document}